\documentclass[aps,prl,reprint,superscriptaddress]{revtex4-1}

\usepackage{graphicx}
\usepackage{dcolumn}
\usepackage{bm}        
\usepackage{amssymb}   
\usepackage{amsmath}
\usepackage{accents}
\usepackage{xcolor}

\begin{document}

\title{Robust Ferroelectricity in Monolayer Group-IV Monochalcogenides}

\author{Ruixiang Fei}
\affiliation{Department of Physics, Washington University in St.
Louis, St. Louis, MO 63130, USA}

\author{Wei Kang*}
\affiliation{HEDPS, Center for Applied Physics and Technology, and
College of Engineering, Peking University, Beijing 100871, China}

\author{Li Yang*}
\affiliation{Department of Physics, Washington University in St.
Louis, St. Louis, MO 63130, USA}

\date{\today}

\begin{abstract}
Ferroelectricity usually fades away when materials are thinned
down below a critical value. Employing the first-principles
density functional theory and modern theory of polarization, we
show that the unique ionic-potential anharmonicity can induce
spontaneous in-plane electrical polarizations and ferroelectricity
in monolayer group-IV monochalcogenides MX (M=Ge, Sn; X=S, Se).
Using Monte Carlo simulations with an effective Hamiltonian
extracted from the parameterized energy space, we show these
materials exhibit a two-dimensional ferroelectric phase transition
that is described by fourth-order Landau theory. We also show the
ferroelectricity in these materials is robust and the
corresponding Curie temperature is higher than room temperature,
making these materials promising for realizing ultrathin
ferroelectric devices of broad interest.
\end{abstract}

\maketitle Ferroelectrics, particularly in the thin-film form that
is most commonly needed for modern devices, is plagued by a
fundamental challenge: the depolarization field - an internal
electric field that competes with, and often destroys,
ferroelectricity \cite{1973firstorder,1994Zhong,2005Scott}. As a
result, the critical thickness in proper ferroelectric materials,
such as perovskite ones, is limited between 12 and 24 $\AA$
\cite{2003Ghosez,2004Fong,2004Rabe-science}. New mechanisms, such
as hyperferroelectrics, are proposed to keep the polarization even
in a single layer of $ABC$ hexagonal semiconductors
\cite{2014Vanderbilt}, but these materials have yet to be
synthesized. Layered van der Waals (vdW) materials may provide
another way to overcome this challenge. For example,
two-dimensional (2D) MoS$_2$ was predicted to be a potentially
ferroelectric material \cite{2014Waghmare}. However, its
ferroelectric $1T$ structure is not thermally stable compared to
the observed $2H$ phase.

A recently reported high-performance thermoelectric material, bulk
SnSe, a group-IV monochalcogenide \cite{2014Zhao}, may give hope
to ferroelectricity because of its giant anharmonic and
anisotropic phonons \cite{2015Hong,2016Tanaka,2014triaxial}, which
are usually the signs of spontaneous symmetry breaking. In
particular, monolayer structures of this MX (M=Ge, Sn; X=S, Se)
family may exhibit giant piezoelectricity \cite{2015Yang,
2015neto} and their electrical polarization displays nonlinear
response to applied strain \cite{2015Yang}. All these clues
motivate us to investigate their potentially spontaneous
polarization and ferroelectricity. Furthermore, ultra-thin
trilayers of these materials have been successfully fabricated
\cite{2013jacs}, making the study of monolayers of immediate
experimental interest. Finally, since dimensionality is a known
factor deciding phase transitions, it is of fundamental interest
to investigate if a ferroelectric phase transition can occur in
these materials and further to find the Curie temperature and
critical phenomena, which may be different from those in bulk
materials
\cite{1973firstorder,1995Sizephase,1994Zhongphase,2014Iniguez}.

\begin{figure}
\centering
\includegraphics[scale=0.42]{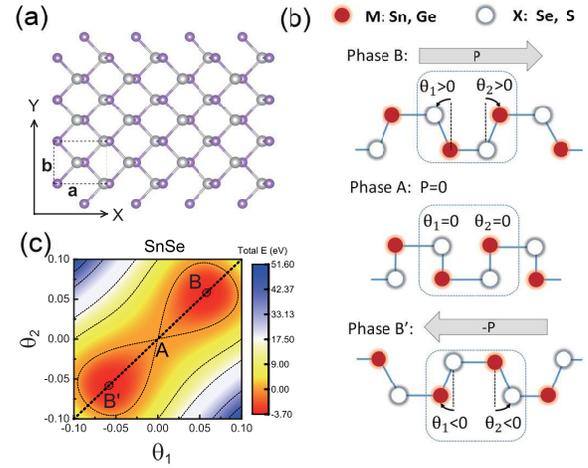}
\caption{\label{FIG. 1} (Color online) (a) Top view of the atomic
structure of monolayer group-IV monochalcogenides. The black line
rectangle is the Brillouin zone, a is the lattice constant along
armchair direction (X), b is that along the zigzag direction (Y).
(b) The schematic side view of the two distorted structures (B and
B') along with the high symmetry phase (A). (c) The DFT-calculated
free-energy surface contour plot of monolayer SnSe.}\label{Fig_1}
\end{figure}

In this work, we show that MX (M=Ge, Sn; X=S, Se) monolayers are a
new family of 2D ferroelectric vdW materials. Using
first-principle calculations \cite{DFT_detail, 1999Vasp, 1996PBE,
2008Phonopy, 1993Vanderbilt, 1994Resta}, we identify two
degenerate polar structures exhibiting spontaneous in-plane
polarization, which avoids the competing depolarization field. The
calculated polarization intensities are similar to those of
typical bulk materials. Moreover, we build an effective
Hamiltonian based on the parameterized energy surface and use it
to investigate the phase transition via Monte Carlo (MC)
simulations \cite{MD_detail}. The calculated Curie temperatures
($T_c$) of these 2D structures are above room temperature, making
them promising for experimental realization and ultra-thin
ferroelectric devices that have been pursued for decades. In
particular, we reveal the relation between $T_c$, configurational
energy barriers, and the spontaneous polarization, demonstrating
that this 2D ferroelectric phase transition obeys a fourth-order
Landau theory. Finally, we calculate a phase diagram for SnSe,
which is essential for material fabrication and measurements.

Bulk MX (M=Ge, Sn; X=S, Se) adopts a layered orthorhombic crystal
structure (space group $Pnma$) at room temperature, which can be
derived from a three-dimensional distortion of the $NaCl$
structure (space group $Cmcm$)\cite{2014Zhao}. Their monolayer
structures keep the $Pnma$ symmetry \cite{Supplement} and the top
view is plotted in Fig. 1 (a). From the side view, we define the
angles $\theta_1$ and $\theta_2$ measured along the x (armchair)
direction shown in Fig. 1 (b) which describes the geometric
distortion. When $\theta_1 = \theta_2 = 0$, the structure converts
to the non-polar $Cmcm$ (phase A) with inversion symmetry, which
is actually the structure of the crystalline insulator materials,
SnTe, PbTe, etc \cite{2015topological,Supplement}. For SnSe, there
are two stable structures which are related by a spatial
inversion, characterized by having both $\theta_1$ and $\theta_2$
positive or both negative. These structures, labeled phases B and
B', are shown in Fig. 1 (b). The free-energy surface for this
class of configurations, obtained via using first-principle
calculations, is presented in Fig. 1 (c), which confirms these
stable structures are connected through a saddle point,
corresponding to phase A. This type of anharmonic double-well
potential strongly hints the existence of ferroelectricity.

\begin{table}
\caption{\label{tab:table1} The ground-state free-energy
(potential barrier) $E_G$($meV$), and the spontaneous polarization
$P_s$ ($10^{-10} C/m$) at zero temperature, and the fitted
parameters in Equation 1 of our studied MXs. A, B, and C are used
to describe the double-well potential. D is the constant
representing the mean-field approximation interaction between
local mode and the nearest neighbors.}
\begin{ruledtabular}
\begin{tabular}{ccccccc}
\multicolumn{6}{ r }{Constants based on DFT calculation} \\
\cline{2-7}
Material    &$E_G$ &$P_s$  & A   &B  &C  &D \\
\hline
SnSe       &-3.758 &1.49  &-5.785  &1.705  &0.317  &10.16 \\
SnS      &-38.30 &2.62 &-19.127   &1.053   &0.275  &8.49\\
GeSe     &-111.99 &3.67  &-15.869  &-3.540   &0.378 &9.74 \\
GeS     &-580.77 &5.06 &-37.822  &-5.422   &0.280  &10.59 \\
\end{tabular}
\end{ruledtabular}
\end{table}

The crucial point is that both B and B' structures are
non-centrosymmetric polar structures. In particular, B and B' can
be transformed into on another by a spatial inversion. Therefore,
if there is a polarization ($P$) in the B phase, the polarization
of the B' phase must be the inverse ($-P$). Our Berry-phase
calculation based on density functional theory (DFT) confirms this
symmetry analysis: these two stable structures (B and B') have
significant spontaneous polarization with opposite polarizing
directions. The polarization values are listed in Table I. If we
estimate the thickness of each layer to be 0.5 nm, the average
bulk value of the polarization in this material is around
$0.3\sim1.0$ $C/m^2$, which is similar to that of traditional
ferroelectric materials such as $BaTiO_3$ and $PZT$ \cite{2004Pan,
2011Xifan,2014Iniguez}.

\begin{figure}
\includegraphics[scale=0.4]{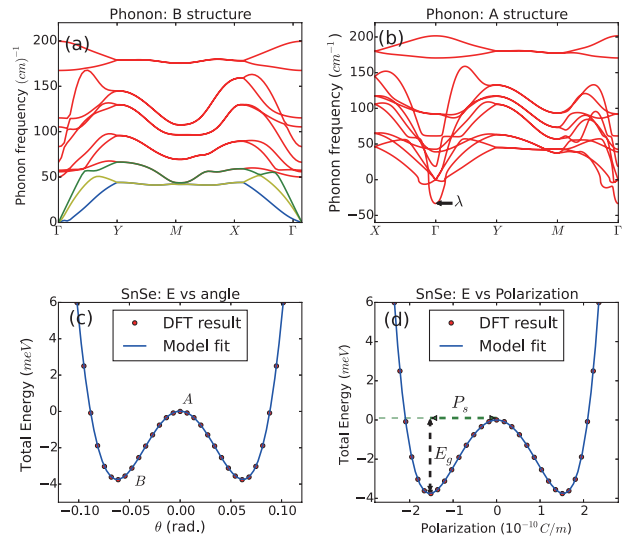}
\caption{(Color online) (a) and (b) are the phonon spectra of the
structures A and B, respectively, which are marked in Fig. 1 (c).
(c) The double-well potential of monolayer SnSe. Red points are
the DFT calculated total energy and the blue line is the model
fit. (d) The double-well potential vs polarization. The $E_G$ is
the ground state energy and $P_s$ is the spontaneous polarization
of the stable phase.}
\end{figure}

In bulk ferroelectrics, the driving mechanism for the symmetry
breaking phases have been widely attributed to soft optical phonon
modes \cite{1968softmode}. These modes correspond to displacive
instabilities in the structure which, below the critical
temperature, freeze the structure in an symmetry-broken phase. At
high temperature, the soft mode will have positive $\omega$, which
will decrease to zero at $T_c$. Below $T_c$, the soft modes have
an imaginary $\omega$. To investigate this possibility in
monolayer SnSe, we plot the phonon dispersion for both the
symmetric phase A and polar phase B (Fig. 2 (a) and 2 (b)).
Clearly there is an imaginary, soft optical mode present which we
call $\lambda$.

Once we have identified soft modes as a likely mechanism for a
phase transition, we would need a way to describe these modes at
finite temperature. Unfortunately, techniques for treating
higher-order anharmonic phonons are currently very limited. One
approximate method of investigating the imaginary modes is to
employ the so-called \emph{renormalization scheme} to calculate
the effective harmonic frequencies at finite temperature
\cite{2016Tanaka}. However, in this scheme, there is no way to
distinguish different dimensionalities, and it may not capture the
essential features of a 2D ferroelectric phase transition.

Rather than try to compute the soft mode directly, we would like
to describe our system in a traditional Landau theory. To do this,
we use as the order parameter the polarization \emph{P} along the
armchair direction, which is the natural choice. Then we map the 2
component ($\theta_1,\theta_2$) free-energy surface (Fig. 1 (c))
to a function of the order parameter $P$. To do this, we observe
that due to the steep gradient of the energy surface in the
perpendicular direction, the structure prefers to stay the so
called \emph{angle-covariant} phase ($\theta_1=\theta_2$), marked
by a dashed line in Fig. 1 (c) \cite{Supplement}. By only
considering this 1D subset of configurations, we greatly simplify
the parameter space of our model.

In Fig. 2 (c), we show the energy along this
\emph{angle-covariant} line, $\theta_1=\theta_2=\theta$ for
monolayer SnSe. Its double-well shape suggests that it may be
described by the known $\phi^4$ potential, which has been widely
used to study spontaneous symmetry breaking. By calculating the
polarization for each value of $\theta$, we can then relate the
free energy $E$ to the polarization $P$.

The potential energy is expressed in the  Landau-Ginzburg
polynomial expansion
\begin{equation}
\begin{split}
&E=\sum\limits_{i}{
\frac{A}{2}(P_i^2)+\frac{B}{4}(P_i^4)+\frac{C}{6}(P_i^6)}+\frac{D}{2}\sum\limits_{<i,j>}{(P_i-P_j)^2}
\end{split}
\end{equation}
which can be viewed as Taylor series, around a reference
structure, of local structural distortions with certain
polarization defined at each cell $P_i$. As shown in Fig. 2 (d),
the first three terms associated with the energy contribution from
the local modes up to the sixth order and they well describe the
anharmonic double-well potential. The last term captures the
coupling between local modes. Compared with the results of
mean-field theory within the nearest-neighbor approximation
(Fig. 3 (a)), the first-principle calculations of supercells
shows that the the coupling is harmonic, confirming the validity
of keeping the second-order dipole-dipole interactions in Eq. (1).
The values of the parameters $A$-$D$ are listed in Table I. Interestingly, the
value for D, describing the average dipole-dipole interaction, is almost same
across these four monolayer materials. This is reasonable because the similar
local structures of these materials.

\begin{figure}
\centering
\includegraphics[scale=0.38]{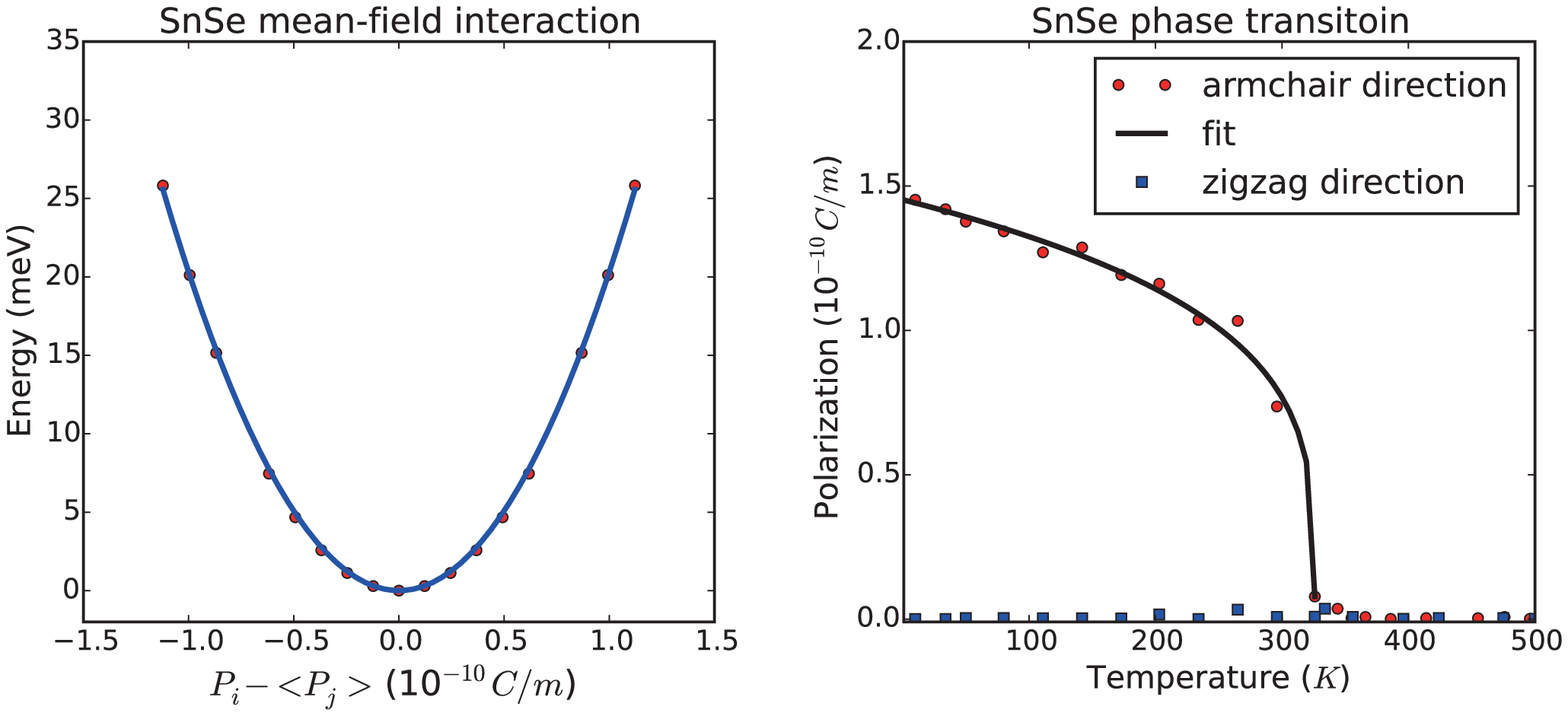}
\caption{(Color online) (a) The dipole-dipole interaction of
monolayer SnSe by using the mean-field theory. The red points are
the DFT-calculated total energy of different $P_i-P_j$. (b)
Temperature dependence of the net polarization obtained from MC
simulations of monolayer SnSe.} \label{Fig_3}
\end{figure}

With this effective Hamiltonian and fitted parameters, we are able
to employ the MC simulation to investigate the phase transition of
monolayer MX. In Fig. 3 (b), we show there is an abrupt transition
at $T_c\approx325$ $K$ for monolayer SnSe. To obtain the critical
exponent and understand the universal critical phenomena, we
employed a fitting procedure that assumes a heuristic form for
$P(T)$:

\begin{equation}
P(T)=\begin{cases}
\mu(T_c-T)^\delta & T<T_c \\
0 & T>T_c
\end{cases}
\end{equation}
where $T_c$ is the Curie temperature, $\delta$ is the critical
exponent, and $\mu$ is a constant prefactor. These fitted results
of monolayer MX are summarized in Table II. The Curie temperature
$T_c$ of monolayer GeS and GeSe are rather large; this is
consistent with their higher energy barriers ($E_G$), hinting that
GeSe and GeS have strong ferroelectric instability. On the other
hand, the smaller $T_c$ of monolayer SnSe and SnS show they have
weak ferroelectric instability, which can be easily tuned by
external field or strain. This is also consistent with our
previously work showing nonlinear polarization response in
strained SnSe and SnS \cite{2015Yang}.

\begin{table}
\caption{\label{tab:table2}Fitted parameters in Eq.2 for MXs.
$T_c$ is the Curie temperature, $\delta$ is the critical exponent,
and $\mu$ is a prefactor.}
\begin{ruledtabular}
\begin{tabular}{cccc}
\multicolumn{3}{ r }{Phase transition} \\
\cline{2-4}
Material    &$T_c$ $(K)$ &$\mu$  & $\delta$   \\
\hline
SnSe       &326 &0.34  &0.25 \\
SnS      &1200 &0.21 &0.35   \\
GeSe     &2300 &0.48 &0.26   \\
GeS     &6400 &0.75 &0.22   \\
\end{tabular}
\end{ruledtabular}
\end{table}

\begin{figure}
\centering
\includegraphics[scale=0.40]{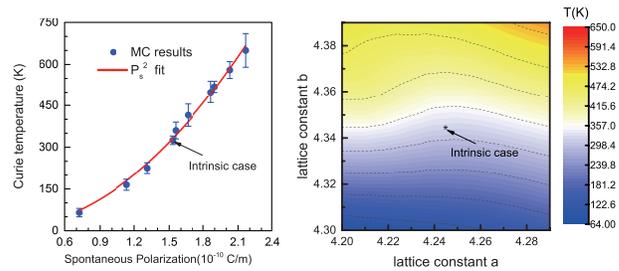}
\caption{(Color online)(a) The relation between Curie temperature
and spontaneous polarization. Blue points including error bars are
Monte Carlo simulation results and the red line is a fit using the
model function described by Eq. 4. (b) The phase diagram of the
ferroelectric phase transition of monolayer SnSe} \label{Fig_4}
\end{figure}

It is important to point out that the Curie temperature $T_c$ may
not be of similar order to the energy barriers. For example, the
barrier of the double-well potential $|E_G|$ of SnSe ($3.758$
$meV$) is much smaller than $k_bT_c$ ($28.02 meV$). This also can
be explained by fourth-order Landau theory, which has been used to
understand the ferroelectricity of perovskites \cite{2004Rappe}.
In this scheme, the free energy can be written as
\begin{equation}
F=\alpha(T-T_c)P^2+\beta P^4
\end{equation}
with $\alpha$,$\beta>0$. The equilibrium polarization is given by
$dF/dP=0$, resulting the Curie temperature
\begin{equation}
T_c=\frac{2\beta} {\alpha} {P_s}^2
\end{equation}

Excitingly, we fit the spontaneous polarization of monolayer SnSe
and find $T_c \sim P_s^2$, which matches perfectly this
fourth-order Landau theory and is the same as traditional
pervskite ferroelectric materials\cite{1968Jamieson,2014Iniguez}.
More precisely, the coefficient, $2\beta/\alpha$, is about $11.09$
$meV/(10^{-10}C/m)^2$. This is very close to the interaction
constant $D$ of monolayer SnSe listed in Table I. Therefore, a
material with weak instability may nevertheless display relatively
high $T_c$ determined by high values of dipole-dipole coupling $D$
and the spontaneous polarization $P_s$.

Finally, since the structures of monolayers are easily affected by
the substrate, fabrication methods and temperature
\cite{2016Salvador}, it is necessary to provide a phase diagram of
their ferroelectricity with different lattice constants.
Therefore, we have varied the two orthogonal lattice constants (a
and b), which can be related to strain (pressure), and calculated
the corresponding Curie temperature $T_c$. As an example, the
phase diagram of monolayer SnSe is presented in Fig. 4 (b).
Interestingly, the ferroelectric transition temperature could be
tuned by a few hundred Kelvins by very small strain (within $\pm1
\%$). This widely tunable range suggests potential challenges for
experimental measurements under different fabricating conditions
and is also promising for the engineering ferroelectricity by
strain.

In conclusion, we predict that monolayer group-IV
monochalcogenides are ferroelectric materials with in-plain
spontaneous polarization. The spontaneous polarization of
monolayer MXs are about $1.49\sim5.06$ $10^{-10}C/m$, and the
Curie temperatures are significantly higher than the energy
barriers between their degenerate ground-state polar structures,
e.g., the $T_c$ of SnSe with weak ferroelectric instability is
around $325K$, although its potential barrier $E_G$ is only 3.758
meV. These properties indicate that monolayer MXs are robust
ferroelectric materials, which could used as the ferroelectric
memory devices. The revealed mechanism of the ferroelectric phase
transition, explained by Landau theory, takes us closer to
understanding the universal critical properties of 2D materials.
Furthermore, the widely-tunable Curie temperature of these
monolayers under strain gives more freedom for engineering
ferroelectric devices.

We acknowledge fruitful discussions with Vy Tran. We are supported
by the National Science Foundation (NSF) CAREER Grant No.
DMR-1455346 and NSF EFRI-2DARE-1542815. The computational
resources have been provided by the Lonestar and Stampede of
Teragrid at the Texas Advanced Computing Center (TACC).

During preparation of this paper, we become aware of a related
theoretical study by P. Hanakata et al. \cite{2016memory}, which
shows that multistability of monolayer SnS and GeSe and the
puckering direction can be switched by application of stress or
electric field.


\begin{thebibliography} {99}

\bibitem{1973firstorder}
I. P. Batra, P. Wurfel, and B. D. Silverman, Phys. Rev. Lett.
\textbf{30}, 384 (1973).

\bibitem{1994Zhong}
W. Zhong, R. D. King-Smith, and D. Vanderbilt, Phys. Rev. Lett.
\textbf{72}, 3618 (1994).

\bibitem{2005Scott}
M. Dawber, K. M. Rabe, and J. F. Scott, Rev. Mod. Phys.
\textbf{77}, 1083 (2005).

\bibitem{2003Ghosez}
J. Junquera and P. Ghosez, Nature \textbf{422}, 506 (2003).

\bibitem{2004Fong}
D. D. Fong, G. B. Stephenson, S. K. Streiffer, J. A. East- man, O.
Auciello, P. H. Fuoss, and C. Thompson, Science \textbf{304}, 1650
(2004).

\bibitem{2004Rabe-science}
C. Ahn, K. Rabe, and J.-M. Triscone, Science \textbf{303}, 488
(2004).

\bibitem{2014Vanderbilt}
K. F. Garrity, K. M. Rabe, and D. Vanderbilt, Phys. Rev. Lett.
\textbf{112}, 127601 (2014).

\bibitem{2014Waghmare}
S. N. Shirodkar and U. V. Waghmare, Phys. Rev. Lett. \textbf{112},
157601 (2014).

\bibitem{2014Zhao}
L.-D. Zhao, S.-H. Lo, Y. Zhang, H. Sun, G. Tan, C. Uher, C.
Wolverton, V. P. Dravid, and M. G. Kanatzidis, Nature
\textbf{508}, 373 (2014).

\bibitem{2015Hong}
C. Li, J. Hong, A. May, D. Bansal, S. Chi, T. Hong, G. Ehlers, and
O. Delaire, Nature Physics \textbf{11}, 1063 (2015).

\bibitem{2016Tanaka}
J. M. Skelton, L. A. Burton, S. C. Parker, A. Walsh, C.- E. Kim,
A. Soon, J. Buckeridge, A. A. Sokol, C. R. A. Catlow, A. Togo, and
I. Tanaka, ArXiv e-prints (2016), arXiv:1602.03762.

\bibitem{2014triaxial}
J. Carrete, N. Mingo, and S. Curtarolo, Applied Physics Letters
\textbf{105}, 101907 (2014).

\bibitem{2015Yang}
R. Fei, W. Li, J. Li, and L. Yang, Applied Physics Letters
\textbf{107}, 173104 (2015).

\bibitem{2015neto}
G. L. C., A. Carvalho, and A. H. Castro Neto, Physical Review B
\textbf{92}, 214103 (2015).

\bibitem{2013jacs}
L. Li, Z. Chen, Y. Hu, X. Wang, T. Zhang, W. Chen, and Q. Wang,
Journal of the American Chemical Society \textbf{135}, 1213
(2013).

\bibitem{1995Sizephase}
S. Chattopadhyay, P. Ayyub, V. R. Palkar, and M. Multani, Phys.
Rev. B \textbf{52}, 13177 (1995).

\bibitem{1994Zhongphase}
W. Zhong, D. Vanderbilt, and K. M. Rabe, Phys. Rev. Lett.
\textbf{73}, 1861 (1994).

\bibitem{2014Iniguez}
J. C. Wojdel and J. Iniguez, Phys. Rev. B \textbf{90}, 014105
(2014).

\bibitem{DFT_detail}
The firrst-principles density functional theory (DFT) were
performed with the Vienna ab initio simulation package (VASP)[20],
using the projector augmented-wave (PAW) method.
Exchange-correlation effects are described by using the
Perdew-Burke-Ernzerhof generalized gradient approximation[21].
electronic wave functions are expanded in a plane-wave basis with
an energy cutoff of 600 eV both for structure relaxation and
charge density. the lattice-dynamics calculations were performed
with the Phonopy [22], using 6x6x1 supercell expansions. The
electronic contribution to the polarization is calculated as a
berry phase using the modern theory of polarization [23, 24].

\bibitem{1999Vasp}
G. Kresse and D. Joubert, Phys. Rev. B \textbf{59}, 1758 (1999).

\bibitem{1996PBE}
J. P. Perdew, K. Burke, and M. Ernzerhof, Phys. Rev. Lett. 77,
3865 (1996).

\bibitem{2008Phonopy}
A. Togo, F. Oba, and I. Tanaka, Phys. Rev. B \textbf{78}, 134106
(2008).

\bibitem{1993Vanderbilt}
R. D. King-Smith and D. Vanderbilt, Phys. Rev. B \textbf{47}, 1651
(1993).

\bibitem{1994Resta}
R. Resta, Rev. Mod. Phys. \textbf{66}, 899 (1994).

\bibitem{MD_detail}
The Monte Carlo (MC) simulations of the effective Hamiltonian in a
periodically repeated box of 15x15x1 unit cells. We adopt the
mean-field theory to describe the interaction between modes. To
get reliable results for bonds displacements, we run at least 80,
000 MC sweeps for finding the thermal-equilibrium state, followed
by, at least, 100, 000 MC sweeps to achieve the thermal averages.
For the temperature close to the transition point, the MC
simulations are run for up to 120, 000 MC sweeps for
thermalization.

\bibitem{Supplement}
See Supplemental material for lattice constants of the stable
phase of monolayer MX. It also includes the $\phi^4$ model to fit
for these four materials, the model analysis for explaining the
angle-covariant case, and the model for geting the phase diagram
of monolayer SnSe.

\bibitem{2015topological}
J. Liu, X. Qian, and L. Fu, Nano letters \textbf{15}, 2657 (2015).

\bibitem{2004Pan}
K. J. Choi, M. Biegalski, Y. L. Li, A. Sharan, J. Schubert, R.
Uecker, P. Reiche, Y. B. Chen, X. Q. Pan, V. Gopalan, L.-Q. Chen,
D. G. Schlom, and C. B. Eom, Science 306, 1005 (2004).

\bibitem{2011Xifan}
X. Wu, K. M. Rabe, and D. Vanderbilt, Phys. Rev. B \textbf{83},
020104 (2011).

\bibitem{1968softmode}
P. A. Fleury, J. F. Scott, and J. M. Worlock, Phys. Rev. Lett.
\textbf{21}, 16 (1968).

\bibitem{2004Rappe}
I. Grinberg and A. M. Rappe, Phys. Rev. B \textbf{70}, 220101
(2004).

\bibitem{1968Jamieson}
S. C. Abrahams, S. K. Kurtz, and P. B. Jamieson, Phys. Rev. 172,
551 (1968).

\bibitem{2016Salvador}
M. Mehboudi, B. M. Fregoso, Y. Yang, W. Zhu, A. van der Zande, J.
Ferrer, L. Bellaiche, P. Kumar, and S. Barraza-Lopez, arXiv
preprint arXiv:1603.03748 (2016).

\bibitem{2016memory}
P. Z. Hanakata, A. Carvalho, D. K. Campbell, and H. S. Park, arXiv
preprint arXiv:1603.00450 (2016).

\end{thebibliography}
\end{document}